\documentclass{sigchi}

\toappear{\scriptsize Permission to make digital or hard copies of all or part of this work for personal or classroom use is granted without fee provided that copies are not made or distributed for profit or commercial advantage and that copies bear this notice and the full citation on the first page. Copyrights for components of this work owned by others than ACM must be honored. Abstracting with credit is permitted. To copy otherwise, or republish, to post on servers or to redistribute to lists, requires prior specific permission and/or a fee. Request permissions from permissions@acm.org. \\
{\emph{CHI 2017, May 6-11, 2017, Denver, CO, USA.} } \\
Copyright \copyright~2017 ACM ISBN 978-1-4503-4655-9/17/05\ ...\$15.00. \\
http://dx.doi.org/10.1145/3025453.3025834}

\usepackage{balance}  
\usepackage{graphics} 
\usepackage[T1]{fontenc}
\usepackage{txfonts}
\usepackage{mathptmx}
\usepackage[pdftex]{hyperref}
\usepackage{color}
\usepackage{booktabs}
\usepackage{textcomp}
\usepackage{bbm}
\usepackage{subfigure}
\usepackage{microtype} 
\usepackage{ccicons}  
\usepackage{float}
\usepackage{todonotes}

\def\plaintitle{Attention Allocation Aid for Visual Search}
\def\plainauthor{Arturo Deza, Jeffrey R. Peters, Grant S. Taylor, Amit Surana, Miguel P. Eckstein}
\def\plainkeywords{Attention; cognitive load; decision making; visual search.}

\makeatletter
\def\url@leostyle{%
  \@ifundefined{selectfont}{
    \def\UrlFont{\sf}
  }{
    \def\UrlFont{\small\bf\ttfamily}
  }}
\makeatother
\def\pprw{8.5in}
\def\pprh{11in}

\definecolor{linkColor}{RGB}{6,125,233}
\hypersetup{%
  pdftitle={\plaintitle},
  pdfauthor={\plainauthor},
  pdfkeywords={\plainkeywords},
  bookmarksnumbered,
  pdfstartview={FitH},
  colorlinks,
  citecolor=black,
  filecolor=black,
  linkcolor=black,
  urlcolor=linkColor,
  breaklinks=true,
}

\begin{document}

\title{\plaintitle}

\numberofauthors{5}
\author{%
  \alignauthor{Arturo Deza\footnotemark\\
    \affaddr{Dynamical Neuroscience}\\
    \affaddr{UC Santa Barbara}\\
    \email{deza@dyns.ucsb.edu}}\\
  \alignauthor{Jeffrey R. Peters}\\
    \affaddr{Mechanical Engineering}\\
    \affaddr{UC Santa Barbara}\\
    \email{jrpeters@engineering.ucsb.edu}\\
  \alignauthor{Grant S. Taylor\\
    \affaddr{U.S. Army}\\
    \affaddr{Aviation Development Directorate, USA}\\
    \email{grant.s.taylor.civ@mail.mil}}\\
  \alignauthor{Amit Surana\\
    \affaddr{United Technologies Research Center, USA}\\
    \email{suranaa@utrc.utc.com}}\\
  \alignauthor{Miguel P. Eckstein\\
\affaddr{Psychological and Brain Sciences, UC Santa Barbara}\\
\email{eckstein@psych.ucsb.edu}}\\
}

\maketitle

\footnotetext{All authors of the paper are also affiliated with the Institute for Collaborative Biotechnologies.}

\begin{abstract}
This paper outlines the development and testing of a novel, feedback-enabled \emph{attention allocation aid (AAAD)}, which uses real-time 
physiological data to improve human performance in a realistic sequential visual search task. Indeed, by optimizing over search duration, the aid improves 
efficiency, while preserving decision accuracy, as the operator identifies and classifies targets within simulated aerial imagery. Specifically, using experimental 
eye-tracking data and measurements about target
detectability across the human visual field, we develop functional models of detection accuracy as a function of search time, number of eye 
movements, scan path, and image clutter. These models are then used by the AAAD in conjunction with real time eye position data to make probabilistic estimations 
of attained search accuracy and to recommend that the observer either move on to the next image or continue exploring the present image. An experimental evaluation in a 
scenario motivated from human supervisory control in surveillance missions confirms the benefits of the AAAD.
\end{abstract}

\category{H.1.2}{[Models and Principles]}{User/Machine Systems--Human factors, Human information processing}
\category{H.4.2}{[Information Systems Applications]}{Types of Systems--Decision Support}
\category{H.m}{[Miscellaneous]}{}
\category{I.6.4}{[Simulation and Modeling]}{Model Validation and Analysis}
\category{}{General Terms}{Experimentation, Human Factors, Verification}

\keywords{\plainkeywords}

\section{Introduction}
The maturation of visual sensor technology has steadily increased the amount of real-time data that
is available in modern surveillance mission scenarios ranging across military, homeland security and commercial applications. In many cases,
it is the job of a human operator to ensure that this data
is processed quickly and accurately. For example, \emph{supervisory systems} involving collaboration between human operators and unmanned
vehicles often require the sequential processing of imagery that is generated by the autonomous vehicles' on-board cameras for the purpose of
finding targets, analyzing terrain, and making key planning decisions~\cite{peters2015human}. The incredible volume of data generated by modern sensors, combined
with the complex nature of modern mission scenarios, makes operators susceptible to information overload and attention allocation 
inefficiencies \cite{cummings2010human}, which can lead to detrimental
performance and potentially dire consequences~\cite{TS-MR:11}. As such, the development of tools to improve human performance in visual data
analysis tasks is crucial to ensuring mission success.


This article focuses on the development and experimental verification of a novel, attention allocation aid that is designed to 
help human operators in a sequential
visual search task, which requires the detection and classification of targets within a simulated landscape. Our study is primarily motivated by
surveillance applications that require humans to find high value targets within videos generated by remote sensors, e.g., mounted on unmanned
vehicles; however, the presented method is applicable to a variety of application domains.

Specifically, the main contribution of this paper is the introduction and experimental verification of a real-time and feedback-enabled \emph{attention allocation aid} (AAAD), 
which 
optimizes the operator's decision speed when  they are engaging in target search, without sacrificing performance.
The motivating observation is that humans have imperfect awareness of the time required to acquire  all task-relevant visual
information during search, and thus are generally inefficient at administering their time when scrutinizing the data sets.
The proposed aid makes real-time automated search duration recommendations based on three key
metrics:
1) visual search time, 2) number of eye movements executed by the observer, and 3) an estimated target detectability
based on prior measurements of both the target visibility across the visual field and the observer's fixations during search. 
In particular, these metrics are used by the aid to estimate the time required for the operator to acquire the visual information that is 
necessary to support the search decisions, and subsequently indicate when this time has elapsed via a simple indicator on the user interface.
We experimentally evaluate the AAAD in a simulated surveillance scenario motivated by human supervisory control, 
and found a factor of $\times1.5$ increase in user efficiency in detecting and classifying targets in realistic visual imagery from a slowly moving sensor. 
The AAAD pipeline is generic and can readily be extended and applied to other sources of images, i.e., satellite images,
astronomical images~\cite{bouman2015computational}, x-rays for medical imaging~\cite{abbey2015approximate,drew2013scanners} or security scanning~\cite{biggs2015improving}, 
and video surveillance~\cite{shanmuga2015eye} which include a human-in-the-loop~\cite{vondrick2011video,vondrick2013efficiently}.

Our rigorous development of the AAAD also includes a number of secondary contributions, including: definition of 
detectability surfaces based on eye-tacking measurements, 
incorporation of image clutter effects, creation of 
a composite exploration map, and utilization of a probabilistic framework for decision making when computing 
overall search satisfaction based on time, eye movements, and detectability scores.

\section{Previous Work}
Research in human computer interaction
has capitalized on basic vision science research
by including samples of the the visual environment through eye movements
(active vision) within a larger modeling framework
that includes cognitive, motor, and perceptual processing
involved in visual search (Tseng and Howes, 2015~\cite{tseng2015adaptation} and 
Halverson and Hofson, 2011~\cite{halverson2011computational} which expands on
the work of Kieras and Meyer, 1997~\cite{meyer1997computational}).
Such models have been proposed for potential use for computer interface
evaluation and design.

Another line of research has focused on how to augment
human capabilities in coordinating multiple tasks. For example,
models have been used to optimize how observers split their
attentional resources when simultaneously conducting two different 
visuo-cognitive tasks~\cite{pavel2003augmented}.

Attention allocation aids have been studied in the context of human
supervisory control of large data acquired by multiple automated
agents (e.g.,~\cite{tbs:92, cummings2010human}). Such scenarios present the challenge of a human having
to inspect large data sets with possible errors
due to visual limitations, attentional bottlenecks, and fatigue.
The use of advanced physiological sensing through eye-tracking technology has become a viable option for both the 
assessment of the operator cognitive state, and the evaluation of operator performance in a number of realistic applications, e.g.~\cite{usaf:10}. One 
line of research attempts to use eye-tracking measurements to detect physiological and cognitive precursors to behavior such as perceived workload, fatigue, or 
situational awareness. Indeed, objective measures such as blink rates~\cite{mevb-lvdh-wn-tv:13}, pupil diameter~\cite{fv-st:14,marshall2002index} , and 
fixation/saccade characteristics~\cite{UA:05}, all have correlations to cognitive processing, although the use of such measurements as reliable 
indicators of operator mental states is not fully understood~\cite{BD-PEP-MLC:09}. If undesirable states can be accurately 
anticipated with physiological measures,
then they can be used to drive 
automated aids that mediate operator resources through, e.g., optimization of task 
schedules~\cite{JRP-LFB:16} or adaptive automation schemes~\cite{dck:14, MS:01}.

Other researchers have utilized eye tracker technology to show increased efficiency of human search by relying on a ``divide and conquer'' 
collaborative search~\cite{zhang2016look,brennan2008coordinating}. In such schemes, multiple observers (usually two) engage in target search
simultaneously with real-time updates of their partners' gaze and access to voice communication.

A novel approach investigated in the current work is the design of an attention allocation aid 
that uses eye-tracking data 
to make real time inferences of the attained target detection accuracy and critically, the time to achieve asymptotic accuracy.
Such estimates, which we will refer to as search satisfaction time, are utilized by the attention allocation aid to recommend that the user
end the current search and move on to the next data set. In addition, if the observer completes search prior to the search satisfaction time,
the eye position data can also be utilized to assess whether some area of the image remains unexplored, and suggest that the observer to further explore that area. 

The success of the proposed approach requires an adequate 
understanding of the relation between fixational eye-movements and the accumulation of sensory evidence supporting task performance.
A critical component to understanding the contribution of eye movement to task performance is the dependence of target detectability with its distance
from the point of fixation, commonly referred to as \emph{retinal eccentricity}~\cite{diaz2012measurements,peli1990contrast,najemnik2005optimal,deza2016can}. 
Indeed, this relationship can be used to build attention-based models for predicting 
performance~\cite{ik-ar:11}. Often, dynamic sensory evidence accumulation models are also dependent upon the nature of the stimuli. 
Our attention allocation aid relies on a set of experiments measuring how accuracy in detecting
the target of interest varies with distance from fixation (retinal eccentricity) and as a function of the presentation time
of the image data. These measurements are then used to implement the AAAD and validate its utility in optimizing search. To our knowledge the current approach for the AAAD
and thorough experimental validation is novel to the field.

We also note that a key difference of our work in comparison to existing literature, is that our attention allocation aid is essentially a back-end search 
optimizer, which tells 
the observer \emph{when} to stop search; rather than advising the observer \emph{where} to 
look (it does not compute fixation cue's or saliency-like maps~\cite{koehler2014saliency,bylinskii2016should}).

\section{Motivation and Hypothesis}
Humans have difficulty assessing when adequate visual 
information has been acquired during challenging search tasks and optimally allocating 
their fixations over different parts of the image~\cite{wolfe2012quit}.
The purpose of the AAAD 
is to utilize in real time the temporal dynamics of the eye-position data and the information acquisition process to recommend to the observers that either all 
information has been acquired and search can be terminated, or further exploration of the image is required. The AAAD is expected to reduce both premature 
image search termination and long periods of image search when no target is present without compromising the search task performance, i.e. detection and false 
alarm rates. Thus, the AAAD should ideally improve observer's efficiency in completing more sequential search tasks in a given allotted time period with 
a level of detection accuracy that is as good or better than search without the AAAD.

In what follows, we present a series of experiments in order to develop, calibrate, and test the effectiveness of the AAAD in a 
visual target search/classification task. Subjects were asked to search for people within simulated aerial images, and subsequently 
classify whether or not the person was holding a weapon. The stimuli and eye-tracking apparatus used in all of our experiments 
are described below.

\begin{figure*}[!t]
\centering
\includegraphics[scale=0.5,clip=true,draft=false,]{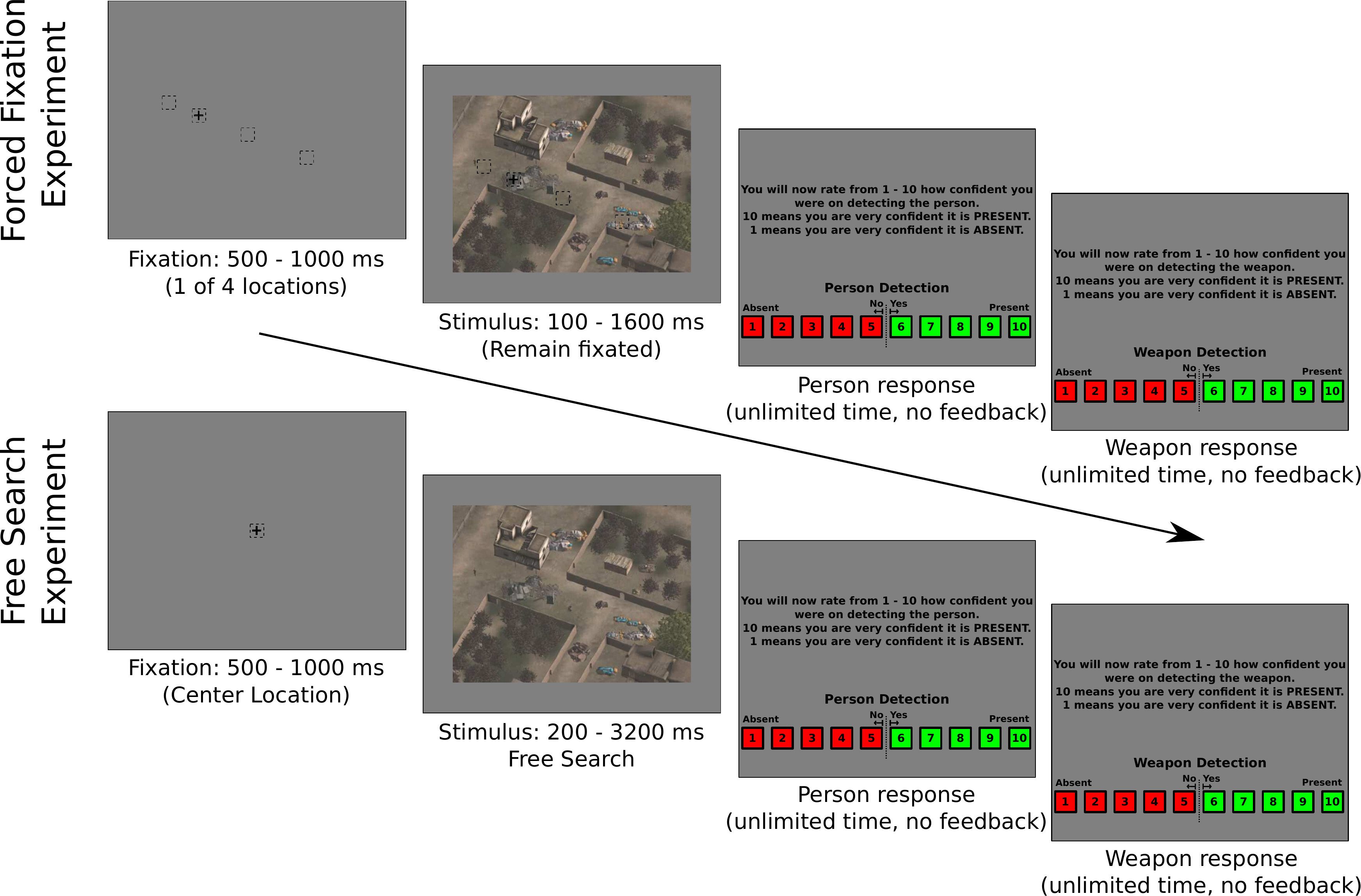}
\caption{Experiment 1: The forced fixation (top) and free search (bottom) experiments to obtain the time, eye movements, and detectability PPCs.
}\label{fig:Experiment_Flow_All}
\end{figure*}

\textbf{Stimuli Creation:} A total of 273 videos were created, each with a total duration of 120 seconds, where a `birds eye' point-of-view camera
rotated slowly around the center. While the video was in a rotating motion, there was no relative motion between any parts of the video.
From a repeated subset of the original 273 videos, a total of $1440$ different short clips were created, 
which were subsequently divided into the $4$ groups (stimuli sets) that were used in subsequent experiments. 
Half of the clips 
had person present, while the other half had person absent.
These short and slowly rotating clips were used instead of still images in our experiment, to simulate 
imagery from a moving sensor in a surveillance scenario.
All clips were shown to participants in a random order. The stimuli used in all our experiments present varying levels of zoom (high, medium, low) and 
clutter (high, medium, low).

\textbf{Apparatus:} An EyeLink 1000 system (SR Research) was used to collect eye-tracking data at a frequency of 
1000 Hz. Each participant sat at a distance of 76 cm from a LCD screen
on gamma display, so that each pixel subtended a visual angle of $0.022\deg/\text{px}$. All video clips were rendered at $1024\times 760$ px 
$(22.5\text{ }\deg\times 16.7\text{ }\deg)$ and a frame rate of 24 fps.
Eye movements with velocity over $22\text{ }\deg/s$ and acceleration over $4000\text{ }\deg/s^2$ were qualified as saccades. Every trial began with a fixation cross, where
each subject had to fixate the cross with a tolerance of $1\text{ }\deg$.

\section{Developing the Attention Allocation Aid (AAAD)}
This section presents details about the development and calibration of the AAAD.
Here, 
preliminary experiments were run to estimate perceptual 
performance curves (PPCs) (describing the relationship between detection accuracy and each of three metrics: time,  
number of eye movements, and detectability) under various levels of image complexity (clutter and zoom levels). 
These PPCs allow the identification of asymptotic detection accuracy, 
which is the primary criterion used by the AAAD to estimate expected detection accuracy in real time.

\subsection{Experiment 1: Psychometric Data Collection}
We performed two preliminary studies to generate the time, eye movement, and detectability PPCs: 
a forced fixation search (no eye movements allowed) and a free search experiment.
The free search data is directly used to compute the relevant PPCs, while the forced fixation data is 
used to calculate detectability surfaces that allow for the computation of the detectability PPC. 
See Figure~\ref{fig:Experiment_Flow_All} for experimental flow.

\subsubsection{Forced Fixation Search}
A total of 13 subjects participated in a \emph{forced fixation search} experiment where the goal was to search within the visual periphery
to identify if there was a person present or absent (yes/no task; $50\%$ probability of person presence) and, in addition, to identify if there was a weapon present or 
absent (yes/no task; $50\%$ probability of weapon presence contingent on person present).
Participants had variable amounts of time (100, 200, 400, 900, 1600 ms) to view each clip. 
Clips were presented in a random order, with the person at a variable 
degree of eccentricity ($1\deg$, $4\deg$, $9\deg$, $15\deg$) from point of fixation. 
Subjects were not made aware of the eccentricity values used in each trial.
They were then prompted with a Likert scale that required them to rate from 1-10 (by clicking on a number) their confidence of person presence. 
A value of 1 indicated strong confidence of person absent, and a value of 10 indicated a strong confidence of person present -- 
intermediate values represented different levels of uncertainty. Values of 1-5 were classified as person absent, and 6-10 were classified as person present. 
A second rating scale (identical to the first) was then presented, requiring the subject to rate their confidence regarding weapon presence.
Participants had unlimited time for making their judgments, although no subject ever took more than 10 seconds per judgment. 
There was no response feedback after each trial. 

Each subject participated in 12 sessions that consisted of 360 clips each.
There were 4 stimuli sets (each set consisted of unique images), and each participants viewed each set 3 times in random order 
without being aware that the images were repeated (4 sets $\times$ 3 times $=$ 12 sessions).
Every set also had the images with aerial viewpoints from different vantage
points (Example: set 1 had the person at 12 o'clock -- as in North, while set 2 had the person at 3 o'clock -- as in East). 
To mitigate fixation bias, all subjects had a unique
fixation point for every trial associated with each particular eccentricity value. All clips were rendered with variable 
levels of clutter.
Each session took approximately one hour to complete. The person,i.e. search target, was of 
size $0.5\text{ }\deg\times 0.5\text{ }\deg$, $1\text{ }\deg\times 1\text{ }\deg$, $1.5\text{ }\deg\times 1.5\text{ }\deg$, 
depending on the zoom level. If a subject fixated outside of a $1\text{ }\deg$ radius around the fixation cross during the trial, 
then the trial was aborted.

\begin{figure}
\centering
\subfigure[Person and weapon detection proportion correct (PC)]{
    \includegraphics[scale=0.27,clip=true,draft=false,]{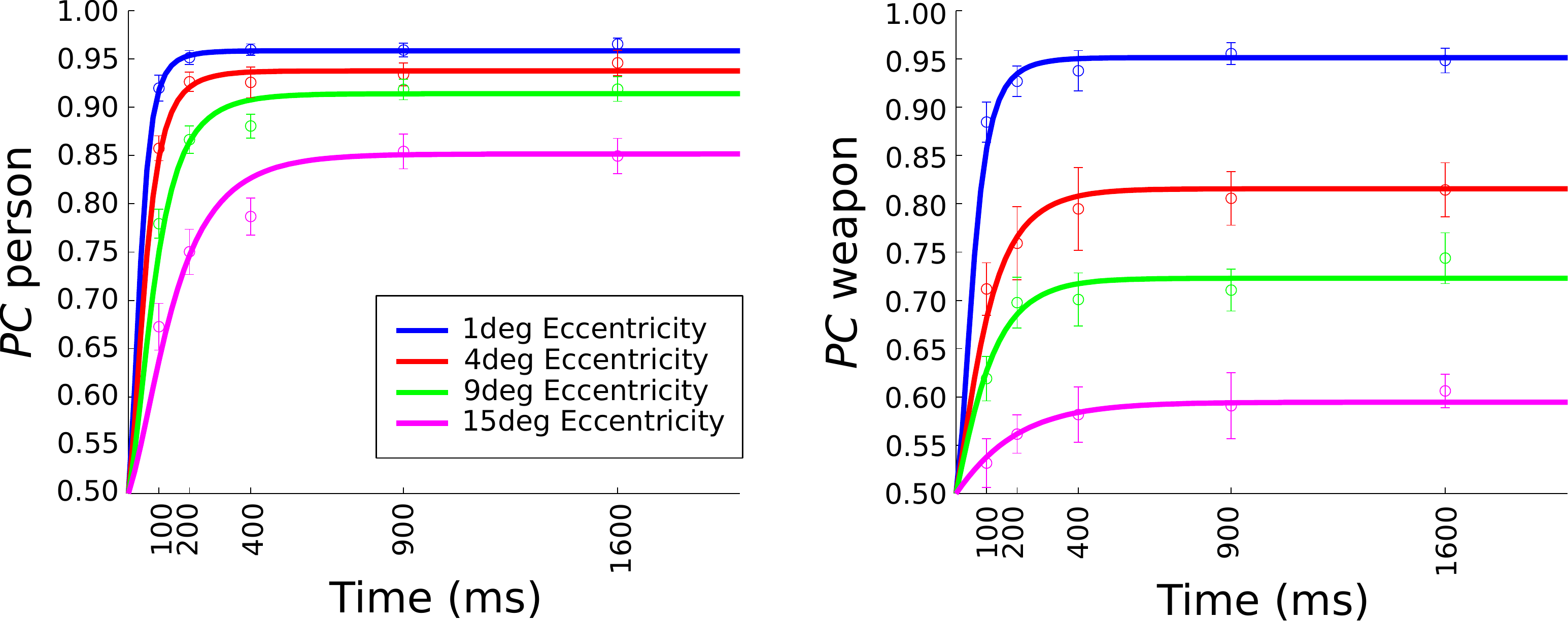}
    \label{fig:Target_Weapon_FF_PC}
}
\subfigure[Person and weapon detectability curves ($d'$)]{
    \includegraphics[scale=0.27,clip=true,draft=false,]{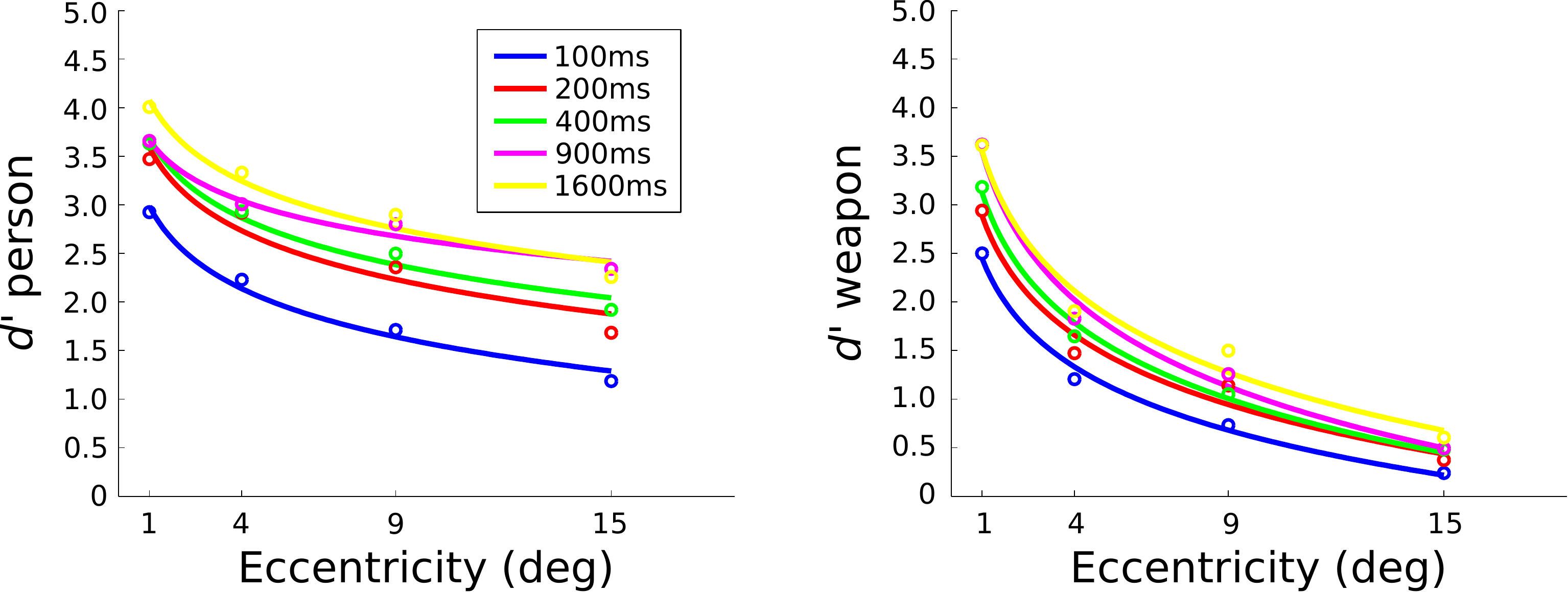}
    \label{fig:Target_Weapon_FF_dprime}
}
\caption{Person and weapon detection performance in proportion correct (PC) and $d'$ space from the forced fixation search experiment. 
Notice that (a) and (b) are dual representations of each other. The bottom curves
in $d'$ space will be used to generate a detectability surface. 
}\label{fig:FF_Data}
\end{figure}

\subsubsection{Free Search}

A total of 11 subjects participated in a \emph{free search} experiment where the goal was to detect and classify the person. Although eye movements were allowed,
subjects were not explicitly told to foveate at the person (although they usually chose to do so). Participants had twice the 
amount (200 ms, 400 ms, 800 ms, 1800 ms, 3200 ms) of time than in the Forced Fixation Search.
All observers began each trial with a fixation at center of the screen.
They then proceeded to scan the scene to find a person and determine if the person was holding a weapon. Once the trial time was over, they were 
prompted with a ``Person Detection'' and ``Weapon Detection'' rating scale, and had to rate from 1-10 by clicking on a number reporting how confident 
they were on detecting/classifying the person. Similar to the forced fixation experiment, participants had unlimited time to make their judgments and did not receive any feedback after each trial. No trials were aborted.

Each subject in the free search experiment participated in 6 sessions that consisted of two sets of 360 unique images. 
In these sessions, each subject viewed one of the two sets of images, and each set was presented 3 times leading to a total of 6 sessions. 
Subjects were not made aware that the sessions were repeated.

\subsection{Fitting Perceptual Performance Curves (PPCs)}
\subsubsection{Motivation of PPCs}
\begin{figure}
\centering
    \includegraphics[width=1.0\columnwidth]{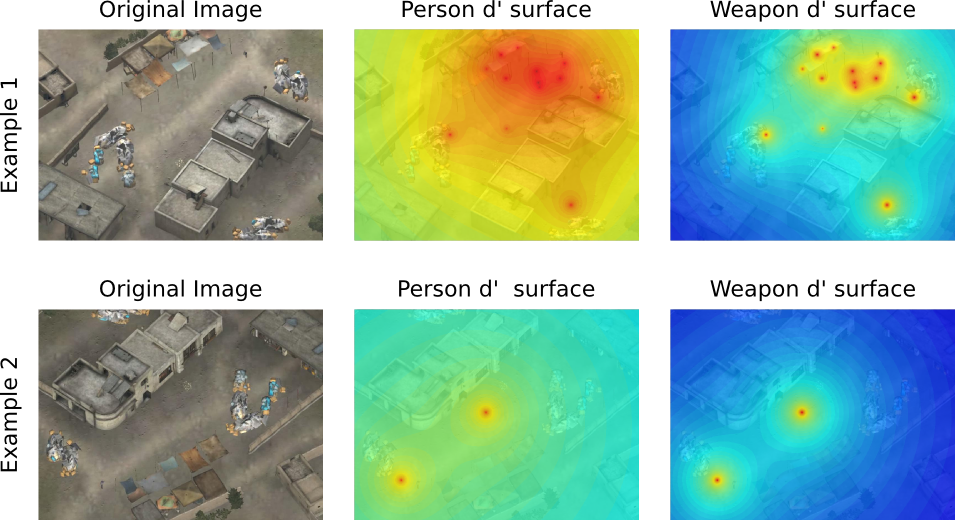}
\caption{Sample person and weapon fixation maps generated from the forced fixation search experiment (Fig.~\ref{fig:Target_Weapon_FF_dprime}). 
These fixation maps are projections of Detectability Surfaces as described in the Supp. Mat. 
}\label{fig:surface_d_prime}
\end{figure}

PPCs were constructed to relate performance to each of three different metrics.
The first metric is visual search time, since it is well known that time affects visual search accuracy~\cite{eckstein2011visual} -- the main intuition 
being that the more time a subject spends scanning an image, the higher the likelihood of detecting the target (person or weapon). The second metric is the number 
of eye movements a subject performs
while engaging in target search. Typically, time will pass on as more eye movements are produced, but there are some cases where scrutiny in classifying or 
detecting a target is needed by spending long periods
of fixation. As an example, one could imagine an \emph{exploitation vs exploration} search scenario where a subject spends 1000 ms on a single fixation,
given the difficulty to \emph{classify} the target (exploitation), as opposed to a scenario where the same subject makes 3 sparse and exploratory 
fixations in the same 1000 ms time window to \emph{find} the target (exploration). For this reason
we chose to make time and eye movements independent metrics for our AAAD system. The third and last metric is detectability. 
Here, a detectability score is constructed by generating a pixel-wise map that quantifies localized information aggregation in different parts of 
the image (as indicated by eye movements), and subsequently combining the result to quantify the target's overall detectability. 
Following our previous example, one could imagine that even if an observer spends the allotted 1000 ms searching for a target
and making $e$ number of eye movements in a small spatial vicinity, it might not be a good strategy compared to spreading fixations across 
the image. See Figure~\ref{fig:surface_d_prime} for an example of such 
fixations overlayed on different images.

We are interested in successful observer detection of the person and the weapon targets. 
Given that our results show that the weapon requires more time to detect than the person, the AAAD recommendation to end search was based on the PPCs
corresponding to the detection of the weapon. Basing the AAAD on the PPCs for the person detection would likely compromise the detection of the weapon.

\begin{figure*}
\centering
\includegraphics[width=2.0\columnwidth]{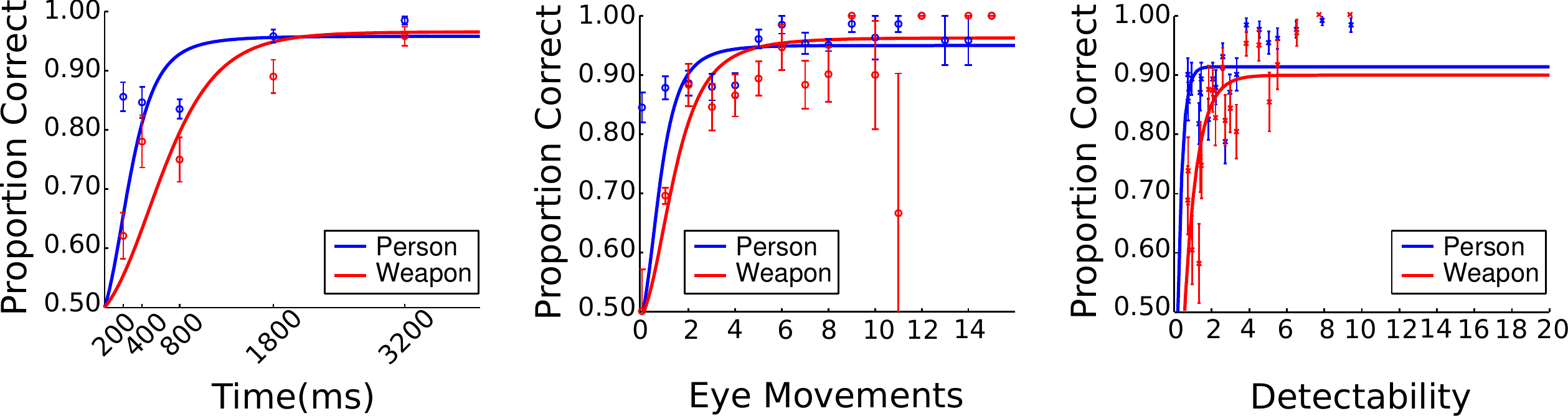}
  \caption{Perceptual Performance Curves (PPCs) for time (left), eye movements (center), 
  and detectability (right) for a (low zoom, high clutter) setting. 
  In general it takes higher time, eye movements, and detectability scores to achieve asymptotic performance for weapon detection \textit{vs} person detection. 
  Also shown are the error bars along each curve. Recall they have been fitted in $d'$ space, and have been re-plotted in PC space (Eq.~\ref{eq:PC_dprime}).}
\label{fig:PPC_curves}
\end{figure*}

\subsubsection{Computing PPCs}
To model the target detection accuracy, we use the observer hit rate (the proportion of trials 
that the observer indicated that a target 
is present, given that a target is actually present in the trial stimuli) 
and false alarm rate (the proportion of trials that the observer indicated 
that the target is present, given that no target is actually pesent in the trial stimuli).
Hit rates and false positive were represented as an empirical detectability 
index ($d'$) and a decision criterion ($\lambda$) using an equal variance normal 
Signal Detection Theory (SDT) model (Green \& Swets)~\cite{green1966signal}. 
We then fit the resulting data with curves to model the functional relationship between the detectability indices and each of the relevant performance metrics.
The best fit functions were then utilized with the equal variance SDT model to generate estimates of attained accuracy in terms of proportion correct (PC).
Notice that proportion correct and hit rate are different since proportion correct takes into account both the hit rate and the correct rejection rate (proportion of trials
in which the observer correctly decided that the target is absent).

For a fixed condition 
and setting (assuming Gaussian signal and noise distributions), the general equations to compute $(d',\lambda)$ are~\cite{wickens2001elementary}:

\begin{equation}
 d' = Z(\text{Hit Rate}) - Z(\text{False Alarm Rate})
\end{equation}
\begin{equation}
\lambda = -Z(\text{False Alarm Rate})
\end{equation}
where $Z(\circ)$ is the inverse of the normal cumulative Gaussian distribution, and the hit/false alarm rates are calculated at the given experimental 
condition and setting.
Consider as an example a condition and setting for the forced fixation search 
experiment: 
Condition = $(4\text{ }\deg,200\text{ ms})$, Setting = $(\text{high zoom},\text{low clutter})$. Likewise, 
a sample condition and setting for the free search experiment: Condition =  $400\text{ ms}$, Setting = $(\text{medium zoom},\text{high clutter})$.

For the Time and Eye-Movements PPCs, we used a straightforward regression to find an exponential relation of 
the form $d'(x) = \alpha(1-e^{-\beta x})$, where $x:$ $\{$Time, Eye Movements$\}$ and $\beta$ is constant, to obtain a continuous 
approximating function for the collection of points $d'$ within a given setting.

To compute final Time and Eye Movements PPC curves, recall that there exists a function $g(\circ)$ that estimates $PC$, 
i.e., $PC(x) = g(d'(x),\lambda(x))$, where:
\begin{equation}\label{eq:PC_dprime}
PC(x) = m(x) (\text{Hit Rate}) + n(x)(1-(\text{False Alarm Rate}))
\end{equation}
and
\begin{equation}
 \text{False Alarm Rate} (x)=Z^{-1}(-\lambda(x))
\end{equation}
\begin{equation}
\text{Hit Rate} (x) = Z^{-1}(d'(x)-\lambda(x))
\end{equation}
where $m(x)$ and $n(x)$ are variables that are contingent on the number of signal (i.e. person/weapon) present 
and signal (i.e. person/weapon) absent trials ($m(x)+n(x)=1,\forall x$), and the hit/false alarm rates are estimates of the true values at $x$.
Here, curves are fitted in $d'$ space, rather than directly from PC space, to deal with possible unbalanced datasets with signal present/absent trials (See Discussion).

In order to obtain the detectability PPC (Figure~\ref{fig:PPC_curves} (right)), we perform a binned regression across all trials between the
respective $PC$ performance from the free viewing task in Experiment 1 and the values of a composite detectability score $D'$ for each image. 
The score $D'$ in each trial was computed by first generating a \emph{detectability surface} from the user's 
fixations and the detectability curves (Figure.~\ref{fig:Target_Weapon_FF_dprime}), and subsequently performing a spatial average. 
We note that the detectability PPC is the only curve that is regressed directly to $PC$ given that the argument of our regression 
function is in a $d'$-like space, thus $\lambda,m(x),n(x)$ need not be calculated. The training data we have for this regression is from the 
free search experiment. 

Further details regarding the generation of the Time PPC, Eye Movements PPC, and Detectability PPC is provided in the Supplementary Material.

\subsection{Performance Criteria and AAAD Functionality}
\label{sec:AAAD_theory}

The AAAD was designed to integrate three different inputs to compute search satisfaction. 
The three previous PPC inputs can be seen as individual metrics on their own,
and are computed independently in the system.

\subsubsection{Search Satisfaction Model}
For computing general search satisfaction we require that all three of the following conditions are simultaneously satisfied:
\begin{equation}
\label{eq:EqT}
Pr[(PC_{max}^T-PC(t))<\epsilon]>\eta,
\end{equation}
\begin{equation}
\label{eq:EqE}
Pr[(PC_{max}^E-PC(e))<\epsilon]>\eta, 
\end{equation}
\begin{equation}
\label{eq:EqD}
Pr[(PC_{max}^D-PC(D'))<\epsilon]>\eta,
\end{equation}
where $PC_{max}^T, PC_{max}^E, PC_{max}^D$ are the (fixed) asymptotic values of $PC$ with respect 
to the time, number of eye-movements, and detectability PPCs given a (zoom, clutter) setting, resp. $PC(t), PC(e), PC(D')$ are the current estimated 
values of $PC$ as calculated by the time, number of eye-movements, and detectability PPCs, respectively, 
and $\epsilon, \eta$ are fixed thresholds.

An image is only said to have been adequately searched if 
\emph{all} criteria~\eqref{eq:EqT},~\eqref{eq:EqE}, and~\eqref{eq:EqD} are simultaneously satisfied. Also 
notice that the criteria~\eqref{eq:EqT},~\eqref{eq:EqE}, and~\eqref{eq:EqD} are all non-decreasing in their respective 
arguments (which are monotonically increasing in time); thus a
condition will never revert to being ``unsatisfied'' after being satisfied. We will refer the above search satisfaction 
criterion as \emph{PC general satisfaction}.

Our motivation for using the above mentioned probabilistic framework is to take into account the error bars of each time, eye movements, and detectability
curves that are zoom and clutter level dependent. We include these error bars as Gaussian standard deviations $\sigma$ in our probabilistic computation:
\begin{equation}
Pr[(PC_{max}-PC)<\epsilon] \rightarrow 1 - Z^{-1}(\frac{PC_{max}-PC-\epsilon}{\sigma}).
\end{equation}
The above strategy can be thought of centering a gaussian $(\mu_x,\sigma_x)\coloneqq(PC,\sigma)$ at every point in the PPC curves, and 
computing how far away the asymptotic performance $PC_{max}$ is from every point in the curve. Thus, we will find and select the minimum point in the curve that fulfills this condition for each time, 
eye movements, and detectability PPC. These are the threshold PPC's that once all of them are reached in real-time on the AAAD system, the AAAD will trigger ``On''. 
A value of $\eta=0.025,\epsilon=0.02$ was selected for our experiments.

\subsubsection{Attention Allocation Aid Design}

\begin{figure*}[!t]
\centering
\includegraphics[width=2.0\columnwidth]{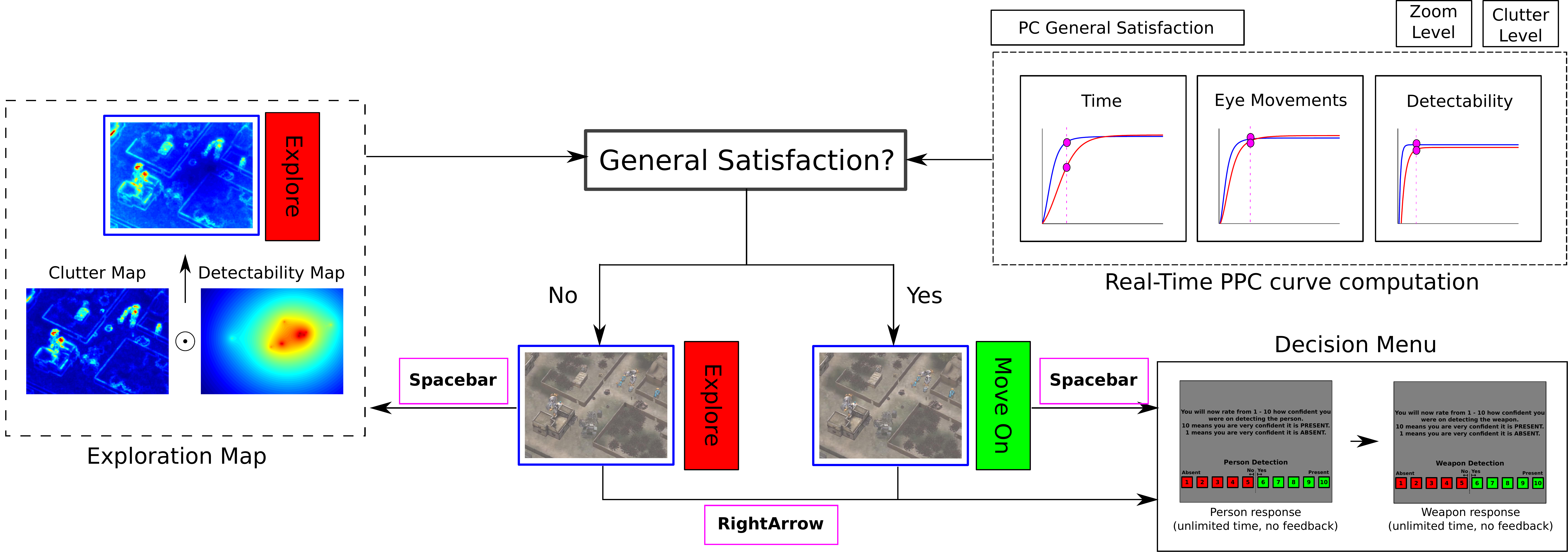}
  \caption{Attention Allocation Aid (AAAD) system diagram. From the start of each trial, the PPC curves are updated in real-time
  in the back-end, waiting for General Search Satisfaction to be achieved. Possible user inputs (highlighted in magenta) are the space bar and the right arrow.
  The right arrow takes the user directly to the decision menu terminating the trial, regardless of Satisfaction. The spacebar will
  lead the observer to the Exploration Map, and later back to the original stimuli if Satisfaction is not achieved, 
  or to the decision menu otherwise}
  \label{fig:AAAD_main_fig}
\end{figure*}

The Attention Allocation Aid system is designed to run in the background of the simulation interface (see Figure \ref{fig:AAAD_main_fig}). The AAAD system starts by computing the clutter and zoom
level of the input image from the slowly evolving video clip in each trial. It is assumed that the level of zoom (high, medium, or low) can be obtained from ground truth settings, given
that a pilot can control a camera's zoom level, while the clutter level can be computed from the input image/video~\cite{yu2013modeling,rosenholtz2007measuring,oliva2004identifying}. For our main experiments, 
we assumed that ground truth was provided to classify images
based on clutter, since our main goal is to prove that the AAAD system works under ideal conditions\footnote{If we did \emph{not} assume \emph{oracle-like} inputs for zoom and clutter levels, then
not finding a significant effect, could be due to poor clutter modeling, rather than poor AAAD system design.}. 
A thorough investigation of the use of clutter models as ground truth predictors is left to future work. 

As the trial progresses, the three PC \textit{vs} $\{$Time, Eye Movements, Detectability$\}$ curves (Figure \ref{fig:PPC_curves}) are updated in 
real-time using gaze location, fixation time, and saccade information obtained from the eye-tracker.
The Time PPC is updated at each frame. The Eye Movements PPC is updated with an interruption-based paradigm -- contingent 
on the eye-tracker detecting an eye movement. The Detectability PPC is updated only after every eye movement event, given that we have the fixation position and time.
Each of the PC Satisfaction conditions (see Eq.~(\ref{eq:EqT},~\ref{eq:EqE},~\ref{eq:EqD})) is monitored independently, and once all criteria are satisfied, the AAAD systems switches from an ``Explore'' to a ``Move On'' state, where observers are encouraged to 
cease search, and make a decision.

Parallel to this, an \emph{Exploration} map is computed in real-time in the back-end. 
The goal of the Exploration map is to inform the searcher where he/she has already searched, and to indicate the highly cluttered regions
where a person is likely to be. The Exploration map has no knowledge of a target present/absent, 
and only uses image clutter and observer fixations.

If an observer attempts to advance to the next image while the AAAD system state is in ``Explore'' state, the Exploration map appears 
for $\sim120$ ms. The map is weighted by previously explored regions 
(computed via the detectability surface; see Supplementary Material), such that,
highly cluttered and non-explored regions are highlighted. The Exploration map is computed as 
follows: $\text{Exploration map} = \text{FC} \odot (1-\text{Detectability Surface})$, where FC is the
\emph{feature congestion}~\cite{rosenholtz2007measuring} dense clutter map, $\odot$ is the element-wise multiplication operator, and 
the detectability surface is normalized to lie in the interval $[0,1]$.

Notice that the only inputs (to the system) that the observer can produce while performing a trial (besides passively providing eye movements), are by pressing the right 
arrow key, which forces
the trial to terminate and the subject to make a decision, irrespective of PC general satisfaction being achieved (for both AAAD, and non-AAAD experimental sessions); or by pressing the
space bar, which activates the Exploration map if PC general satisfaction is not achieved, and terminates the trial if otherwise.

\begin{figure*}
\centering
\includegraphics[width=2.0\columnwidth]{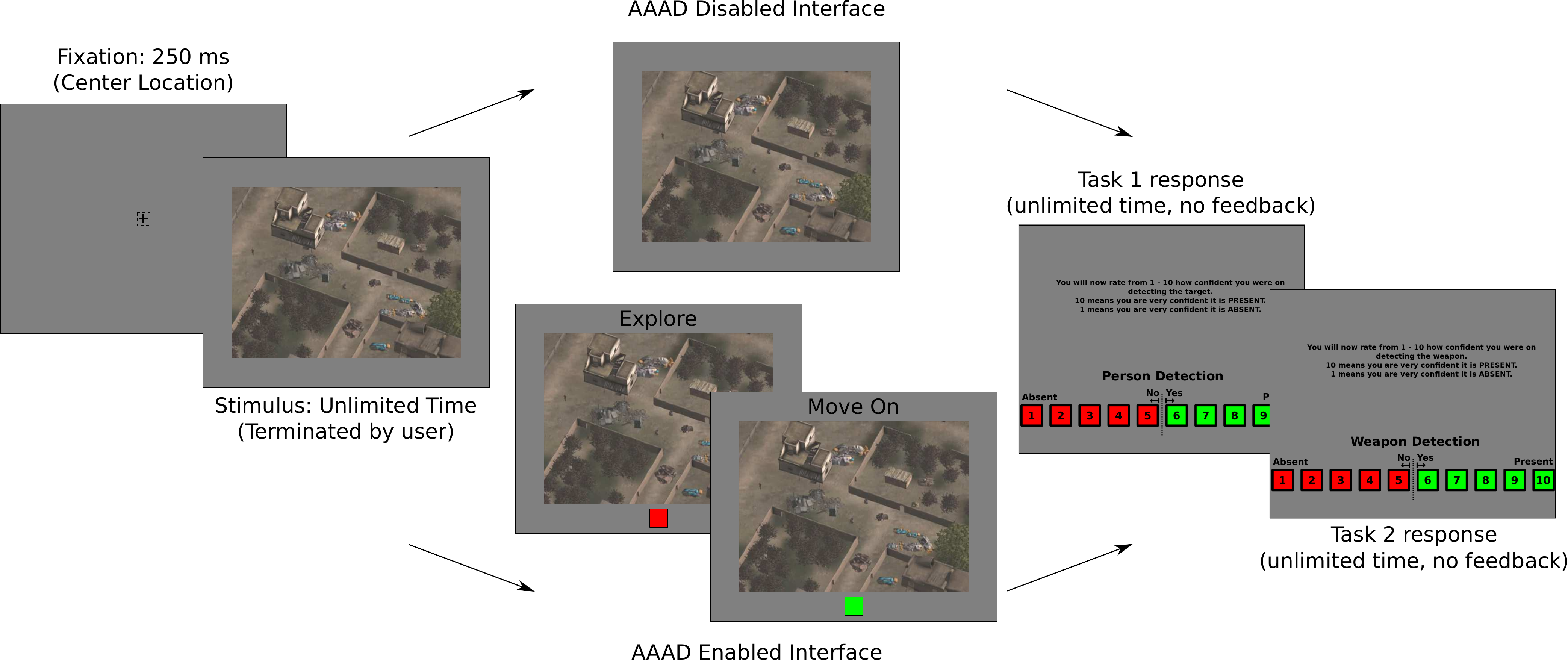}
  \caption{Experimental flow for Experiment 2 where we test the Attention Allocation Aid (AAAD) for condition 1 (AAAD off; top) and condition 2 (AAAD on; bottom). The top stream 
  illustrates the non-AAAD condition, while the bottom stream illustrates
  the AAAD condition. Notice that the text and color in the AAAD enabled interface are coupled together: ``Explore'' = red, ``Move On'' = green. The red and green 
  colors in the decision menus are independent of the AAAD colors.}
\label{fig:AAAD_Flow}
\end{figure*}

\section{Evaluating the Attention Allocation Aid}
In this section we summarize a second experiment used to evaluate the effectiveness of the AAAD. Two experimental conditions were considered: 
A person and weapon search and classification experiment with and without AAAD.
The goal of this experiment is to objectively measure any improvements in the search task performance when the subjects are 
assisted by the AAAD.
Since it was the first time for all of our second group of subjects to participate in an eye tracking experiment, we decided to add two additional practice sessions 
(twenty minutes each) where we would verbally explain the non-AAAD and AAAD system. 

After the practice sessions, half of the subjects were tested starting with the AAAD condition and the other half started without the AAAD condition. We counterbalanced our 
participants to reduce possible learning effects. The group of participants
involved in Experiment 2, did not participate in and were not aware of Experiment 1. 
The completion of the first 2 practice sessions plus Experiment 2 with both conditions (counterbalanced)
took an estimate of 2 hours for each subject. The same person and weapon present/absent statistics of Experiment 1 were used for Experiments 2 and 3.
Figure~\ref{fig:AAAD_Flow} illustrates the design of the search task without (top stream) and with (bottom stream) the AAAD.

\subsection{Experiment 2: AAAD Evaluation}
\subsubsection{Condition 1: Target search without AAAD}
\label{Sec:FF_Search}
A total of 18 subjects 
participated in a non-AAAD target search experiment where the goal was to complete
as many trials as possible in a 20 minute interval without sacrificing task performance, where the task per trial was to correctly
detect and classify the target in the minimum amount of time.

Target detection involved reporting if the person was present or absent in the scene, and target classification involved 
reporting whether or not the person had a weapon. Although eye movements were allowed, subjects were not told explicitly to foveate at the target
(although this was usually the case).
In other words, it was possible for the subject to move on to the next trial by detecting the target in the periphery~\cite{deza2016can}.
A fixation cross was placed at the center of the screen for uniform starting conditions across participants. 
After terminating search, subjects were prompted with a ``Person Detection'' rating 
scale where they had to rate their confidence in a person's presence on a scale from 1-10 by clicking
on a number. A ``Weapon Detection'' rating scale then appeared where subjects also had to rate
their weapon detection confidence from a scale from 1-10. Participants had unlimited time for making their 
judgments, though no subject took more than 10 seconds per judgment.
There was no response feedback, i.e., whether their detection responses were correct after each trial.

\subsubsection{Condition 2: Target search with AAAD}
The same 18 subjects participated in a target search experiment in presence of the AAAD where the goal was same as in the previous condition.
In this experiment, the AAAD was visibly turned on for the participants.
They saw a text message above the center stimuli with a caption: ``Explore'' or ``Move On'', and there was a colored square below the stimuli that was 
colored red or green
depending on the AAAD status. Participants were told to think of the AAAD as a stoplight: when it was red they should keep looking for the person/weapon, and should only move on to the next trial if the light turned green 
or if they were confident that they had either found the person/weapon or there was no person/weapon present.

\begin{table}
  \centering
  \small
  \begin{tabular}{l r r}
    & \multicolumn{2}{c}{\small{\textbf{System Evaluation}}} \\
    \cmidrule(r){2-3}
    & \multicolumn{1}{c}{\small\textit{Non-AAAD}} & \multicolumn{1}{c}{\small\textbf{AAAD}} \\
    \midrule
    Average Trial Number (\#) & $54.33\pm2.26$ & $\mathbf{57.94\pm1.74}$ \\
    \midrule
    Average Mean Time per Trial (s) & $2.88\pm0.42$ & $\mathbf{1.94\pm0.18}$\\
    \midrule
    Person Hit Rate ($\%$) & $96.39\pm1.42$ & $\mathbf{96.94\pm0.88}$ \\
    Weapon Hit Rate ($\%$) & $89.37\pm2.92$ & $\mathbf{92.35\pm2.30}$ \\
    Person False Alarm Rate ($\%$) & $2.88\pm2.63$ & $\mathbf{2.42\pm2.04}$\\
    Weapon False Alarm Rate ($\%$) & $8.19\pm5.54$& $\mathbf{7.15\pm4.70}$ \\
    Person Miss Rate ($\%$) & $3.61\pm1.42$ & $\mathbf{3.06\pm0.88}$ \\
    Weapon Miss Rate ($\%$) & $10.63\pm2.92$ & $\mathbf{7.65\pm2.30}$ \\
    Person Correct Rejection Rate ($\%$) & $97.12\pm2.63$ & $\mathbf{97.58\pm2.04}$\\
    Weapon Correct Rejection Rate ($\%$) & $91.81\pm5.54$ & $\mathbf{92.85\pm4.70}$ \\
    \midrule
    Mean Trial time \textit{vs} Time Trigger (s) & $2.32\pm0.10$ & $\mathbf{1.37\pm0.05}$ \\
    Mean Trial time \textit{vs} EyeMvmt Trigger (s) & $2.35\pm0.10$ & $\mathbf{1.37\pm0.05}$ \\
    Mean Trial time \textit{vs} Detect. Trigger (s) & $2.70\pm0.13$ & $\mathbf{1.10\pm0.06}$ \\    
    Mean Trial time \textit{vs} General Trigger (s) & $2.70\pm0.13$ & $\mathbf{1.10\pm0.06}$ \\
    \midrule
  \end{tabular}
  \caption{General results of the systems evaluation without and with AAAD. It should be noted that subjects were counterbalanced (half-split) to start with or without the AAAD during evaluation
  to compensate for learning effects. Average refers to the mean computed across observers. 
  }
   ~\label{tab:table1}
  \vspace{-15pt}
\end{table}

\section{Results}
Table~\ref{tab:table1} summarizes the results of both conditions in the second experiment. 
There is a significant increase in number of trials per person ($M=3.61$, $SD=5.69$, $t(17) = 2.813$, $p=0.015$, two-tailed), 
as well as a significant decrease in mean trial time ($M=-0.31$, $SD=1.36$, $t(17) = -2.613$, $p=0.009$ ,two-tailed) between AAAD conditions.
Overall performance is stable, though slight improvement is seen with the AAAD. Significant differences of 
trial \emph{vs} trigger times are also found. The AAAD was ran in the back-end (but not visible to the observers) in the non-AAAD condition, to compute
these measures. 
Notice that there is a virtual speed-up factor of: $\times 1.5$, in terms of average mean time per trial across observers when using the AAAD.

In addition we performed a related samples t-test (for person $t_P$ and weapon $t_W$ detection) between the 
hit rates ($M_P=0.54$, $SD_P=3.65$, $t_P(17)=0.497$, $p=0.626$, two-tailed; $M_W=2.97$, $SD_W=12.42$, $t_W(17)=1.016$, $p=0.324$, two-tailed), 
false alarm rates ($M_P=-0.46$, $SD_P=2.98$, $t_P(17)=-0.655$, $p=0.521$, two-tailed; $M_W=-1.03$, $SD_W=6.42$, $t_W(17)=-0.684$, $p=0.503$, two-tailed), 
misses ($M_P=-0.54$, $SD_P=4.65$, $t_P(17)=-0.497$, $p=0.625$, two-tailed; $M_W=-2.97$, $SD_W=12.42$, $t_W(17)=-1.016$, $p=0.323$, two-tailed), 
and correct rejections ($M_P=0.46$, $SD_P=2.98$, $t_P(17)=0.655$, $p=0.521$, two-tailed; $M_W=1.03$, $SD_W=6.42$, $t_W(17)=0.684$, $p=0.503$, two-tailed), 
and found no significant differences between non-AAAD and AAAD conditions. This last finding is somewhat ideal as the AAAD is intended to either preserve
or improve these measures.

Finally, we decided to compare the trigger times of the time, eye movements, detectability, and general satisfaction conditions
of the non-AAAD sessions with the AAAD sessions. 
For comparison, we subtract the final trial time minus the respective trigger time. As such, these times can be thought of as offsets.
Note that although the non-AAAD condition does not show any visible assistant to the observer, the PPCs 
are still being computed in the back-end for the purposes of comparative data analysis.
We performed four independent samples t-tests (Time: $t_T$, Eye Movements: $t_{EM}$, Detectability: $t_D$, General: $t_G$) 
collapsing all trials across all observers 
for these times and found significant differences for all trigger case scenarios, supporting the utility of the AAAD:
$t_T(1920)=-8.46$, $p<0.0001$, two-tailed; $t_{EM}(1900)=-9.03$, $p<0.0001$, two-tailed; $t_{D}(1192)=-11.52$, $p<0.0001$, two-tailed; $t_G(1190)=-11.56$, $p<0.0001$, two-tailed.

%
%

\section{Discussion}
\textbf{Extensions of model validity beyond current scenarios:} Our results show the potential of a new approach in attention 
allocation aids that optimizes human search performance by utilizing real time fixational eye movements
with prior measurements of target visibility across the visual field and as a function of time. However, there are various potential questions 
about the generalization of the model
across search scenarios.

Our development of the AAAD assumed a target that is present
in 50\% of the images. A logical question arises as to whether the framework can generalize
to real scenarios in which the target is present less frequently. Our model
fits herein are performed using signal detection theory metrics 
(Green and Swets, 1967~\cite{green1966signal}) that partition performance into an index
of detectability, which is invariant to target prevalence, 
and a decision criterion, which has an optimal value (maximizing proportion 
correct) that varies with target prevalence. The model curves, which are
utilized to make recommendations to the user, specify proportion correct
as a function of time, eye movements, etc. and will vary with target
prevalence. However, the model can generalize such curves to varying estimated
target prevalence assuming an optimal decision criterion for the given
prevalence. Although, we have not tested the generalization experimentally,
the theory accommodates such scenarios and generalizations.

For simplicity, our current work used a single target when developing the AAAD. 
Another natural question to ask is whether the proposed AAAD can still be used if 
there is the possibility of multiple targets within a given image. Indeed, it could be the case that multiple targets could change
how the aid operates within a given application. 
In some multiple-target scenarios where the detection of \emph{even one} target is sufficient to trigger a decision, 
our strategy may still apply.
For example, for some medical applications, such as screening mammography, finding at least one
suspicious target triggers a follow-up diagnostic mammogram. In other
applications, localization of each individual targets is important and might 
require additional development of multiple target model curves to use in conjunction with a prior
distribution of the number of targets within the images.

\textbf{Impact of computer vision developments on proposed AAAD framework:} The recent
advanced in computer vision might seem to diminish the contributions of the proposed scheme if one assumes that all human search 
will eventually be
replaced by machines. This is yet another reasonable question to ask, since vision has thrived in recent years, in part due to significant advances of 
Deep Learning~\cite{lecun2015deep,krizhevsky2012imagenet}. State of the art object recognition 
algorithms~\cite{simonyan2014very,ren2015faster,he2015deep} have achieved 
high performance on certain datasets (MNIST~\cite{lecun1998gradient}, CIFAR~\cite{krizhevsky2009learning}, ImageNet~\cite{russakovsky2015imagenet}).
However, the images in 
these datasets typically present ideal scenes with large objects at the center of an image 
 and, currently, the ability of state of the art algorithms
to find small or occluded objects in cluttered scenes (MSCOCO~\cite{lin2014microsoft}) remains well-below that of humans. Moreover, computers often show glaring errors that humans
would not make in what have been called \emph{adversarial} examples~\cite{goodfellow2014explaining} in the computer vision 
community (e.g. by rigging individual pixel values in an image which `hacks' a classifier,
a computer can wrongly predict that a white noise-like image is a school bus with 99\% confidence~\cite{nguyen2015deep}).
Furthermore, there is still a fundamental lack of understanding with regard to the effects of computer-aided detection aids as a substitute for human observers in many application domains. For example, 
computer automated detection is prevalent in some countries
to flag potential locations for radiologists scrutinizing x-ray mammograms. Yet there is no consensus about its contributions to improving radiologists'
diagnosis accuracy (e.g. Eadie, Taylor, \& Gibson, 2012~\cite{eadie2012systematic}). As a result of these deficiencies, human observation is still heavily 
relied upon in a number of applications. As a result, there are many ongoing efforts
to reduce errors and optimize human visual search in life-critical tasks from military surveillance~\cite{ALSA2006}, to security 
baggage screening~\cite{hale2015designing},
and medical imaging~\cite{alexander2010reducing}.

\begin{figure*}
\centering
\subfigure[Sample Person Present and Weapon Present Stimuli.]{
    \includegraphics[scale=0.33,clip=true,draft=false,]{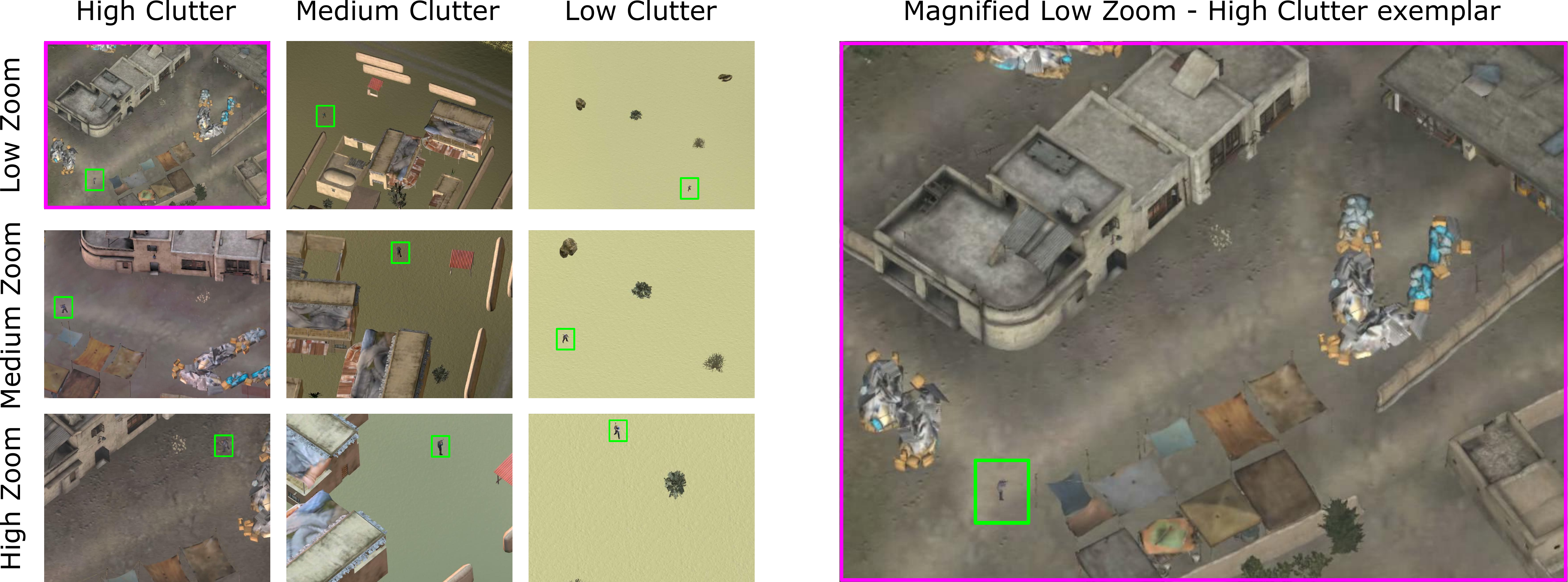}
    \label{fig:Stimuli_TPWP}
}
\caption{Sample Stimuli of Experiments 1 and 2. Left: we show random samples of person present, and weapon 
present in multiple clutter and 
zoom conditions. Right: we show a magnified version of the (Low Zoom, High Clutter) setting.
The box in green has been overlayed on each subimage to reveal the location 
of the person and weapon when applicable.
}\label{fig:Total_Stimuli}
\end{figure*}

What is quickly becoming prevalent across many applications is the use of a computer aid that assists humans in localizing potential 
targets~\cite{giger2014medical}. The proposed
AAAD framework does not take into account the presence of a computer aid flagging potential target locations. In some cases, the presence of a computer aid is known to guide search
with the risk of leading to over-shortened searches and missed targets that are not flagged by the computer 
aid~\cite{lyell2016automation}. The underlying model in the proposed
AAAD allows calculation of an estimated observer accuracy given a pattern of fixations, time and the target visibility across the visual field. In principle,
the model could be used to predict if an observer is short-cutting  their search (due to the presence of the computer aid) and to alert the observer to further
search the image/s. Thus, the developed AAAD framework could be potentially integrated with a computer aid, although its main contribution would likely shift
from reducing search times to reducing missed targets.

\textbf{Potential contribution beyond current application:} Although
the presented work introduces an AAAD within the context of a very specific task
and images, our work serves as a proof of concept for a decision aid design approach that can potentially be applied to a variety 
of other applications including baggage screening and medical imaging. The model within the AAAD predicts
performance on any given trial as a function of time and pattern of fixations and could be
potentially used for quantifying the probability that a target was missed on a given image given the observers' search pattern.
Such probabilities of error could be stored with the images and used later to identify images that require secondary
inspection by an additional $2^{\text{nd}}$ human observer.

Arguably, the main limitation of the AAAD is that the model relies on empirically measured curves describing the 
detectability of the target across the visual field and as a function of time. We are currently investigating
how to predict target detectability by analyzing image properties such as clutter in real time, which 
would greatly benefit the application
of the model to broader domains.

\section{Conclusion}
Our experiments show evidence that our real-time enabled support decision system dubbed \emph{AAAD} 
optimizes user efficiency in terms of an increase in the number of trials done as well as a decrease in time spent per each trial, while maintaining performance
such as target hit rate and false alarm rate.
Thus, the AAAD system has successfully integrated asymptotic performance of search time, eye movements and target 
detectability. We have described how to fully implement such system through an initial psychometric experiments
to find perceptual performance curves for target search, as well as two consequent experiments that verify the benefits of the AAAD. Future computer-human interaction based 
systems could benefit from implementing AAAD-like systems where having a human-in-the-loop is critical to finding a target even beyond surveillance 
systems, e.g., medical imaging, astronomical data imagery and remote sensing.
\vspace{-5pt}
\section{Acknowledgements}
This work has been sponsored by the U.S. Army Research Office and the Regents of the University of California, through Contract Number W911NF-09-D-0001 for 
the Institute for Collaborative Biotechnologies, and that the content of the information does not necessarily reflect the position or the policy of the Government 
or the Regents of the University of California, and no official endorsement should be inferred.
\balance{}

\bibliographystyle{SIGCHI-Reference-Format}
\bibliography{chi2016}


\begin{thebibliography}{00}


\ifx \showCODEN    \undefined \def \showCODEN     #1{\unskip}     \fi
\ifx \showDOI      \undefined \def \showDOI       #1{{\tt DOI:}\penalty0{#1}\ }
  \fi
\ifx \showISBNx    \undefined \def \showISBNx     #1{\unskip}     \fi
\ifx \showISBNxiii \undefined \def \showISBNxiii  #1{\unskip}     \fi
\ifx \showISSN     \undefined \def \showISSN      #1{\unskip}     \fi
\ifx \showLCCN     \undefined \def \showLCCN      #1{\unskip}     \fi
\ifx \shownote     \undefined \def \shownote      #1{#1}          \fi
\ifx \showarticletitle \undefined \def \showarticletitle #1{#1}   \fi
\ifx \showURL      \undefined \def \showURL       #1{#1}          \fi

\bibitem{abbey2015approximate}
{Craig~K. Abbey}, {Frank~W. Samuelson}, {Adam Wunderlich}, {Lucretiu~M.
  Popescu}, {Miguel~P. Eckstein}, {and} {John~M. Boone}. 2015.
\newblock \showarticletitle{Approximate maximum likelihood estimation of
  scanning observer templates}. In {\em SPIE Medical Imaging}. International
  Society for Optics and Photonics, 94160O--94160O.
\newblock


\bibitem{UA:05}
{Ulf Ahlstrom}. 2005.
\newblock \showarticletitle{Subjective workload ratings and eye movement
  activity measures}.
\newblock  (2005).
\newblock


\bibitem{ALSA2006}
{Sea Applications~Center Air, Land}. 2006.
\newblock \showarticletitle{FM 3-04.15, Tactics, Techniques, and Procedures for
  the Tactical Employment of Unmanned Aircraft Systems (UAS)}.
\newblock  (2006).
\newblock


\bibitem{alexander2010reducing}
{Kate Alexander}. 2010.
\newblock \showarticletitle{Reducing error in radiographic interpretation}.
\newblock {\em The Canadian Veterinary Journal\/} {51}, 5 (2010), 533.
\newblock


\bibitem{biggs2015improving}
{Adam~T. Biggs} {and} {Stephen~R. Mitroff}. 2015.
\newblock \showarticletitle{Improving the efficacy of security screening tasks:
  A review of visual search challenges and ways to mitigate their adverse
  effects}.
\newblock {\em Applied Cognitive Psychology\/} {29}, 1 (2015), 142--148.
\newblock


\bibitem{bouman2015computational}
{Katherine~L. Bouman}, {Michael~D. Johnson}, {Daniel Zoran}, {Vincent~L Fish},
  {Sheperd~S Doeleman}, {and} {William~T Freeman}. 2015.
\newblock \showarticletitle{Computational Imaging for VLBI Image
  Reconstruction}.
\newblock {\em arXiv preprint arXiv:1512.01413\/} (2015).
\newblock


\bibitem{brennan2008coordinating}
{Susan~E Brennan}, {Xin Chen}, {Christopher~A Dickinson}, {Mark~B Neider},
  {and} {Gregory~J Zelinsky}. 2008.
\newblock \showarticletitle{Coordinating cognition: The costs and benefits of
  shared gaze during collaborative search}.
\newblock {\em Cognition\/} {106}, 3 (2008), 1465--1477.
\newblock


\bibitem{bylinskii2016should}
{Zoya Bylinskii}, {Adri{\`a} Recasens}, {Ali Borji}, {Aude Oliva}, {Antonio
  Torralba}, {and} {Fr{\'e}do Durand}. 2016.
\newblock \showarticletitle{Where should saliency models look next?}. In {\em
  European Conference on Computer Vision}. Springer, 809--824.
\newblock


\bibitem{cummings2010human}
{Mary~L. Cummings}, {Sylvain Bruni}, {and} {Paul~J. Mitchell}. 2010.
\newblock \showarticletitle{Human supervisory control challenges in
  network-centric operations}.
\newblock {\em Reviews of Human Factors and Ergonomics\/} {6}, 1 (2010),
  34--78.
\newblock


\bibitem{deza2016can}
{Arturo Deza} {and} {Miguel~P. Eckstein}. 2016.
\newblock \showarticletitle{Can Peripheral Representations Improve Clutter
  Metrics on Complex Scenes?}. In {\em Neural Information Processing Systems}.
\newblock


\bibitem{diaz2012measurements}
{Ivan Diaz}, {Miguel~P Eckstein}, {Ana{\"\i}s Luyet}, {Pierre Bize}, {and}
  {Fran{\c{c}}ois~O Bochud}. 2012.
\newblock \showarticletitle{Measurements of the detectability of hepatic
  hypovascular metastases as a function of retinal eccentricity in CT images}.
  In {\em SPIE Medical Imaging}. International Society for Optics and
  Photonics, 83180J--83180J.
\newblock


\bibitem{BD-PEP-MLC:09}
{Birsen Donmez}, {Patricia~E Pina}, {and} {ML Cummings}. 2009.
\newblock \showarticletitle{Evaluation criteria for human-automation
  performance metrics}.
\newblock In {\em Performance Evaluation and Benchmarking of Intelligent
  Systems}. Springer, 21--40.
\newblock


\bibitem{drew2013scanners}
{Trafton Drew}, {Melissa Le-Hoa Vo}, {Alex Olwal}, {Francine Jacobson},
  {Steven~E Seltzer}, {and} {Jeremy~M Wolfe}. 2013.
\newblock \showarticletitle{Scanners and drillers: Characterizing expert visual
  search through volumetric images}.
\newblock {\em Journal of vision\/} {13}, 10 (2013), 3--3.
\newblock


\bibitem{eadie2012systematic}
{Leila~H Eadie}, {Paul Taylor}, {and} {Adam~P Gibson}. 2012.
\newblock \showarticletitle{A systematic review of computer-assisted diagnosis
  in diagnostic cancer imaging}.
\newblock {\em European journal of radiology\/} {81}, 1 (2012), e70--e76.
\newblock


\bibitem{eckstein2011visual}
{Miguel~P. Eckstein}. 2011.
\newblock \showarticletitle{Visual search: A retrospective}.
\newblock {\em Journal of Vision\/} {11}, 5 (2011), 14--14.
\newblock


\bibitem{giger2014medical}
{Maryellen~L Giger}. 2014.
\newblock \showarticletitle{Medical imaging and computers in the diagnosis of
  breast cancer}. In {\em SPIE Optical Engineering+ Applications}.
  International Society for Optics and Photonics, 918908--918908.
\newblock


\bibitem{goodfellow2014explaining}
{Ian~J Goodfellow}, {Jonathon Shlens}, {and} {Christian Szegedy}. 2014.
\newblock \showarticletitle{Explaining and harnessing adversarial examples}.
\newblock {\em arXiv preprint arXiv:1412.6572\/} (2014).
\newblock


\bibitem{green1966signal}
{DM Green} {and} {JA Swets}. 1966.
\newblock \showarticletitle{Signal detection theory and psychophysics. 1966}.
\newblock {\em New York\/}  {888} (1966), 889.
\newblock


\bibitem{hale2015designing}
{Kelly~S Hale}, {Katie Del~Giudice}, {Jesse Flint}, {Darren~P Wilson},
  {Katherine Muse}, {and} {Bonnie Kudrick}. 2015.
\newblock \showarticletitle{Designing, developing, and validating an adaptive
  visual search training platform}. In {\em International Conference on
  Augmented Cognition}. Springer, 735--744.
\newblock


\bibitem{halverson2011computational}
{Tim Halverson} {and} {Anthony~J Hornof}. 2011.
\newblock \showarticletitle{A computational model of “active vision” for
  visual search in human--computer interaction}.
\newblock {\em Human--Computer Interaction\/} {26}, 4 (2011), 285--314.
\newblock


\bibitem{he2015deep}
{Kaiming He}, {Xiangyu Zhang}, {Shaoqing Ren}, {and} {Jian Sun}. 2015.
\newblock \showarticletitle{Deep residual learning for image recognition}.
\newblock {\em arXiv preprint arXiv:1512.03385\/} (2015).
\newblock


\bibitem{dck:14}
{David~C. Klein}. 2014.
\newblock {\em Using Adaptive Automation to Increase Operator Performance and
  Decrease Stress in a Satellite Operations Environment}.
\newblock Ph.D. Dissertation. Colorado Technical University.
\newblock


\bibitem{koehler2014saliency}
{Kathryn Koehler}, {Fei Guo}, {Sheng Zhang}, {and} {Miguel~P Eckstein}. 2014.
\newblock \showarticletitle{What do saliency models predict?}
\newblock {\em Journal of vision\/} {14}, 3 (2014), 14--14.
\newblock


\bibitem{ik-ar:11}
{Ian Krajbich} {and} {Antonio Rangel}. 2011.
\newblock \showarticletitle{Multialternative drift-diffusion model predicts the
  relationship between visual fixations and choice in value-based decisions}.
\newblock  {108}, 33 (2011), 13852--13857.
\newblock


\bibitem{krizhevsky2009learning}
{Alex Krizhevsky}. 2009.
\newblock \showarticletitle{Learning multiple layers of features from tiny
  images}.
\newblock  (2009).
\newblock


\bibitem{krizhevsky2012imagenet}
{Alex Krizhevsky}, {Ilya Sutskever}, {and} {Geoffrey~E Hinton}. 2012.
\newblock \showarticletitle{Imagenet classification with deep convolutional
  neural networks}. In {\em Advances in neural information processing systems}.
  1097--1105.
\newblock


\bibitem{lecun2015deep}
{Yann LeCun}, {Yoshua Bengio}, {and} {Geoffrey Hinton}. 2015.
\newblock \showarticletitle{Deep learning}.
\newblock {\em Nature\/} {521}, 7553 (2015), 436--444.
\newblock


\bibitem{lecun1998gradient}
{Yann LeCun}, {L{\'e}on Bottou}, {Yoshua Bengio}, {and} {Patrick Haffner}.
  1998.
\newblock \showarticletitle{Gradient-based learning applied to document
  recognition}.
\newblock {\it Proc. IEEE} {86}, 11 (1998), 2278--2324.
\newblock


\bibitem{lin2014microsoft}
{Tsung-Yi Lin}, {Michael Maire}, {Serge Belongie}, {James Hays}, {Pietro
  Perona}, {Deva Ramanan}, {Piotr Doll{\'a}r}, {and} {C~Lawrence Zitnick}.
  2014.
\newblock \showarticletitle{Microsoft coco: Common objects in context}. In {\em
  European Conference on Computer Vision}. Springer, 740--755.
\newblock


\bibitem{lyell2016automation}
{David Lyell} {and} {Enrico Coiera}. 2016.
\newblock \showarticletitle{Automation bias and verification complexity: a
  systematic review}.
\newblock {\em Journal of the American Medical Informatics Association\/}
  (2016), ocw105.
\newblock


\bibitem{marshall2002index}
{Sandra~P. Marshall}. 2002.
\newblock \showarticletitle{The index of cognitive activity: Measuring
  cognitive workload}. In {\em Human factors and power plants, 2002.
  proceedings of the 2002 IEEE 7th conference on}. IEEE, 7--5.
\newblock


\bibitem{meyer1997computational}
{David~E Meyer} {and} {David~E Kieras}. 1997.
\newblock \showarticletitle{A computational theory of executive cognitive
  processes and multiple-task performance: Part I. Basic mechanisms.}
\newblock {\em Psychological review\/} {104}, 1 (1997), 3.
\newblock


\bibitem{najemnik2005optimal}
{Jiri Najemnik} {and} {Wilson~S. Geisler}. 2005.
\newblock \showarticletitle{Optimal eye movement strategies in visual search}.
\newblock {\em Nature\/} {434}, 7031 (2005), 387--391.
\newblock


\bibitem{nguyen2015deep}
{Anh Nguyen}, {Jason Yosinski}, {and} {Jeff Clune}. 2015.
\newblock \showarticletitle{Deep neural networks are easily fooled: High
  confidence predictions for unrecognizable images}. In {\em 2015 IEEE
  Conference on Computer Vision and Pattern Recognition (CVPR)}. IEEE,
  427--436.
\newblock


\bibitem{oliva2004identifying}
{Aude Oliva}, {Michael~L. Mack}, {Mochan Shrestha}, {and} {Angela Peeper}.
  2004.
\newblock \showarticletitle{Identifying the perceptual dimensions of visual
  complexity of scenes}.
\newblock


\bibitem{pavel2003augmented}
{Misha Pavel}, {Guoping Wang}, {and} {Kehai Li}. 2003.
\newblock \showarticletitle{Augmented cognition: Allocation of attention}. In
  {\em System Sciences, 2003. Proceedings of the 36th Annual Hawaii
  International Conference on}. IEEE, 6--pp.
\newblock


\bibitem{peli1990contrast}
{Eli Peli}. 1990.
\newblock \showarticletitle{Contrast in complex images}.
\newblock {\em JOSA A\/} {7}, 10 (1990), 2032--2040.
\newblock


\bibitem{JRP-LFB:16}
{Jeffrey~R. Peters} {and} {Luca~F. Bertuccelli}. 2016.
\newblock \showarticletitle{Robust Task Scheduling for Multi-Operator
  Supervisory Control Missions}.
\newblock {\em AIAA Journal on Aerospace Information Systems\/} (2016).
\newblock
\newblock
\shownote{To Appear.}


\bibitem{peters2015human}
{Jeffrey~R. Peters}, {Vaibhav Srivastava}, {Grant~S. Taylor}, {Amit Surana},
  {Miguel~P. Eckstein}, {and} {Francesco Bullo}. 2015.
\newblock \showarticletitle{Human supervisory control of robotic teams:
  integrating cognitive modeling with engineering design}.
\newblock {\em IEEE Control Systems\/} {35}, 6 (2015), 57--80.
\newblock


\bibitem{ren2015faster}
{Shaoqing Ren}, {Kaiming He}, {Ross Girshick}, {and} {Jian Sun}. 2015.
\newblock \showarticletitle{Faster R-CNN: Towards real-time object detection
  with region proposal networks}. In {\em Advances in neural information
  processing systems}. 91--99.
\newblock


\bibitem{rosenholtz2007measuring}
{Ruth Rosenholtz}, {Yuanzhen Li}, {and} {Lisa Nakano}. 2007.
\newblock \showarticletitle{Measuring visual clutter}.
\newblock {\em Journal of vision\/} {7}, 2 (2007), 17--17.
\newblock


\bibitem{russakovsky2015imagenet}
{Olga Russakovsky}, {Jia Deng}, {Hao Su}, {Jonathan Krause}, {Sanjeev
  Satheesh}, {Sean Ma}, {Zhiheng Huang}, {Andrej Karpathy}, {Aditya Khosla},
  {Michael Bernstein}, {and} {others}. 2015.
\newblock \showarticletitle{Imagenet large scale visual recognition challenge}.
\newblock {\em International Journal of Computer Vision\/} {115}, 3 (2015),
  211--252.
\newblock


\bibitem{MS:01}
{Mark~W. Scerbo}. 2001.
\newblock \showarticletitle{Adaptive Automation}.
\newblock In {\em Neuroergonomics: {T}he Brain At Work}, {Raja Parasuraman}
  {and} {Matthew Rizzo} (Eds.). 239--252.
\newblock


\bibitem{TS-MR:11}
{Thom Shanker} {and} {Matt Richtel}. 2011.
\newblock In New Military, Data Overload Can Be Deadly.
\newblock The New York Times.   (January 16, 2011).
\newblock


\bibitem{shanmuga2015eye}
{Karthikeyan Shanmuga~Vadivel}, {Thuyen Ngo}, {Miguel Eckstein}, {and} {B.S.
  Manjunath}. 2015.
\newblock \showarticletitle{Eye tracking assisted extraction of attentionally
  important objects from videos}. In {\em Proceedings of the IEEE Conference on
  Computer Vision and Pattern Recognition}. 3241--3250.
\newblock


\bibitem{tbs:92}
{Thomas~B. Sheridan}. 1992.
\newblock {\em Telerobotics, Automation, and Human Supervisory Control}.
\newblock MIT press.
\newblock
\showISBNx{9780262515474}


\bibitem{simonyan2014very}
{Karen Simonyan} {and} {Andrew Zisserman}. 2014.
\newblock \showarticletitle{Very deep convolutional networks for large-scale
  image recognition}.
\newblock {\em arXiv preprint arXiv:1409.1556\/} (2014).
\newblock


\bibitem{tseng2015adaptation}
{Yuan-Chi Tseng} {and} {Andrew Howes}. 2015.
\newblock \showarticletitle{The adaptation of visual search to utility, ecology
  and design}.
\newblock {\em International Journal of Human-Computer Studies\/}  {80} (2015),
  45--55.
\newblock


\bibitem{usaf:10}
{{US Air Force}}. 2010.
\newblock {\em Report on technology horizons, a vision for {Air Force Science
  And Technology} during 2010--2030}.
\newblock {T}echnical {R}eport. AF/ST-TR-10-01-PR, United States Air Force.
  Retrieved from http://www.af.mil/shared/media/document/AFD-100727-053.pdf.
\newblock


\bibitem{fv-st:14}
{Fran{\c c}ois Vachon} {and} {S\'ebastien Tremblay}. 2014.
\newblock \showarticletitle{What Eye Tracking Can Reveal about Dynamic
  Decision-Making}. In {\em Int.\ Conference on Applied Human Factors and
  Ergonomics}. Krak\'ow, Poland, 3820--3828.
\newblock


\bibitem{mevb-lvdh-wn-tv:13}
{Marlies~E. van Bochove}, {Lise Van~der Haegen}, {Wim Notebaert}, {and} {Tom
  Verguts}. 2013.
\newblock \showarticletitle{Blinking predicts enhanced cognitive control}.
\newblock {\em Cognitive, Affective, \& Behavioral Neuroscience\/} {13}, 2
  (2013), 346--354.
\newblock


\bibitem{vondrick2013efficiently}
{Carl Vondrick}, {Donald Patterson}, {and} {Deva Ramanan}. 2013.
\newblock \showarticletitle{Efficiently scaling up crowdsourced video
  annotation}.
\newblock {\em International Journal of Computer Vision\/} {101}, 1 (2013),
  184--204.
\newblock


\bibitem{vondrick2011video}
{Carl Vondrick} {and} {Deva Ramanan}. 2011.
\newblock \showarticletitle{Video Annotation and Tracking with Active
  Learning}. In {\em NIPS}.
\newblock


\bibitem{wickens2001elementary}
{Thomas~D. Wickens}. 2001.
\newblock {\em Elementary signal detection theory}.
\newblock Oxford university press.
\newblock


\bibitem{wolfe2012quit}
{Jeremy~M. Wolfe}. 2012.
\newblock \showarticletitle{When do I quit? The search termination problem in
  visual search}.
\newblock In {\em The influence of attention, learning, and motivation on
  visual search}. Springer, 183--208.
\newblock


\bibitem{yu2013modeling}
{Chen-Ping Yu}, {Wen-Yu Hua}, {Dimitris Samaras}, {and} {Greg Zelinsky}. 2013.
\newblock \showarticletitle{Modeling clutter perception using parametric
  proto-object partitioning}. In {\em Advances in Neural Information Processing
  Systems}. 118--126.
\newblock


\bibitem{zhang2016look}
{Yanxia Zhang}, {Ken Pfeuffer}, {Ming~Ki Chong}, {Jason Alexander}, {Andreas
  Bulling}, {and} {Hans Gellersen}. 2016.
\newblock \showarticletitle{Look together: using gaze for assisting co-located
  collaborative search}.
\newblock {\em Personal and Ubiquitous Computing\/} (2016), 1--14.
\newblock


\end{thebibliography}
\balance{}
\newpage

\section{Supplementary Material}
\subsection{Additional PPC Computation Details}
\textbf{Time PPC}: We use the equal-variance assumption Gaussian model to retrieve $\lambda$, s.t. $\lambda(x)=-Z(\bar{f})$, where
$\bar{f}$ is the average number of false alarms across all 5 time conditions (200 ms, 400 ms, 800 ms, 1800 ms, 3200 ms). We use this model because there
is an equal number of person present/absent, and weapon present/absent trials (contingent on person present). This implies: $m(x)=n(x)=0.5$.

\textbf{Eye Movements PPC}: Eye movements were quantized in all our experiments as the number of saccades.
We estimated the observer bias $\lambda$, for every $x=0,1,2,3,...,15$ saccades, and later computed a weighted average (inversely proportional
to the error bar size) obtaining a single estimate $\lambda_0$ to serve over all $x$ conditions.
We approximated $m(x)=m_0,n(x)=n_0$, with the constants being proportional to the
average number of trials present and absent across eye movement conditions.

\textbf{Detectability PPC}: To create a composite detectability score ($D'$) used as input of our Detectability PPC 
(Figure~\ref{fig:PPC_curves}, right), 
we created a \emph{detectability surface} as seen in Figure~\ref{fig:surface_d_prime} based on the forced fixation detection curves. First, 
the forced fixation detectability curves as shown in (Fig.~\ref{fig:Target_Weapon_FF_dprime}) were obtained 
from the forced fixation experimental data in the dual $d'$ space as a function of eccentricity $e$ and parameterized by search time. 
A logarithmic fit of the form $d'(e) = \alpha + \beta \log(e)$, where $\alpha, \beta$ are constants, was used to 
produce the curve for each search time condition. Detectability was offset by $1\text{ }\deg$ of eccentricity given 
the forced fixation tolerance during Experiment 1, 
and to avoid interpolation errors at $d'(0\text{ }\deg)$.

Then, we use the curves of Fig.~\ref{fig:FF_Data} to create a detectability surface as follows.  We start by collecting all $j$ fixation point locations 
and times: $(z_j,t_j)$ and creating a pixel-wise mesh $\textbf{I}$ of possible person eccentricities across the 
image on a per observer basis. We compute the $d'$ value using the curves in Fig.~\ref{fig:FF_Data} for every 
point of the mesh $\mathbf{I}$, given the time each fixation took and the distance between the fixation 
location and the mesh-point location, i.e., $t=t_j$ and $z=||z_j-z_{(p, q)}|| \forall (p,q)\in \textbf{I}$.
Note that any fixation time and eccentricity can be extrapolated from the forced fixation experiment.
The surface is produced by putting a normal axis to the image plane at the location of fixation $j$, and 
performing a 3D rotation around this axis. This idea is fundamentally an adaptation of the concept of \textit{surface of revolutions}, where the generatrix
is the forced fixation function $d'(z)$ (Fig.~\ref{fig:FF_Data}), and the axis of rotation is perpendicular from image $\mathbf{I}$ at location $z_j$.
We refer to this as a \emph{single-fixation surface}, which we denote $(\text{Detectability Surface})_j$.
Notice that this procedure does not require knowledge of the person location, and can be thought of a non-normalized probability map that shows likelihood of 
finding a person on the image given any fixation location and time.
The previous computation can be easily vectorized. 

Each generated surface is added linearly over each observer fixation $j$ to 
compute the (multiple-fixation) detectability surface: $\text{Detectability Surface} = \sum_j (\text{Detectability Surface})_j$ over 
the image $\mathbf{I}$. We define the final composite
detectability score $D'$ as the spatial mean of the detectability surface over the image $\textbf{I}$. 
These $d'$ scores can be added with an $L\infty$-norm or max (single-look strategy), L1-norm (late-variability model), and L2-norm 
(likelihood ratio observer)~\cite{wickens2001elementary}. We use a L1-norm since real-time
computations of detectability surfaces are facilitated through vectorized addition: $O(n)$ \textit{vs} $O(n^2)$. 

\begin{figure*}
\centering
\includegraphics[width=2.0\columnwidth]{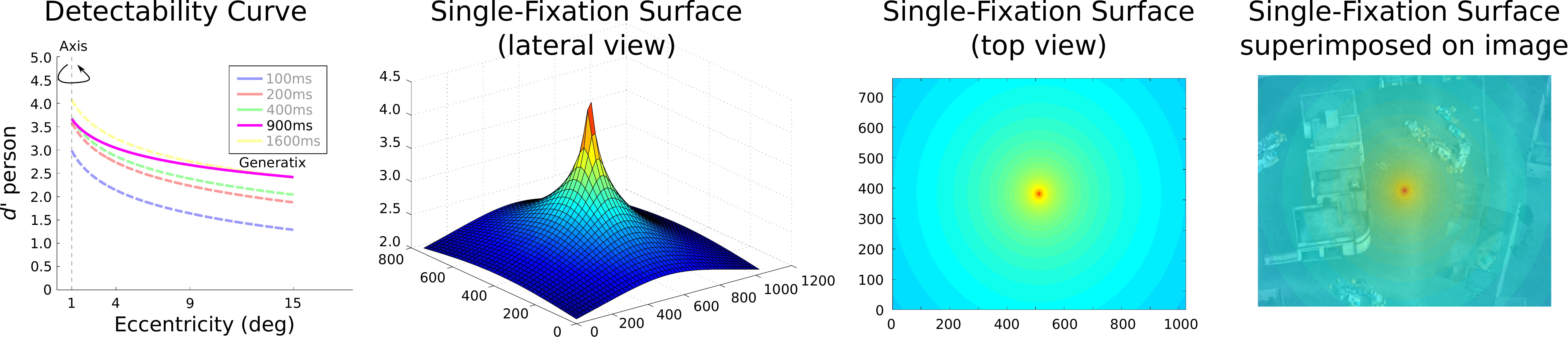}
  \caption{$D'$ score generation: In this example, the observer made a single 900 ms fixation. We use the forced fixation search experiment detectability surface to get the logarithmic
  function that represents the decay in target detectability to compute a surface of revolutions, with the generatrix as the magenta curve, and the axis 
  always at $1\deg$ on the curve, but centered at the fixation location (center of the image, in this case). The Surface is then projected in 2D on top of the image, and the final $D'$
  detectability score (used in the Detectability PPC) is the mean of this projected surface.}
\label{fig:Supp_Surface}
\end{figure*}

\subsection{Instruction Delivery}
Delivery of instructions to participants seemed to play a role in the experiment and the use of the AAAD. In preliminary versions of the AAAD,
some subjects mistakenly thought that the AAAD light (red/green) was a sign of whether the person/weapon was present or absent. 
In other words, they thought that the AAAD's goal was to tell them if the person/weapon was present or absent, instead of thinking of the AAAD as a search time indicator
on when to terminate search. Instructions delivered in the final version (those used for the experiments herein), made this distinction clear.

The most emphasized sentence of the instructions for Experiments 2 and 3 
was: \textit{``Observers should strive to accomplish as many trials as possible without sacrificing detection performance''}. 
In addition, subjects were informally told: \textit{``(...) you want to do the trials fast, but you don't want to rush and end up making careless mistakes''}.

Other details that should be taken into account for functionality of the AAAD is the possibility of subjects that were very conservative \emph{i.e.} 
they only moved on when the AAAD
triggered on; or, in contrast, subjects that ignored the AAAD in general \emph{i.e.} they rarely followed the AAAD given reasons such as curiosity, possibility of deception,
or general slow response times. While we cannot control for this type of behavior, this was not seen in our results subjects pool, and analyzing the data
over 18 subjects is a sufficient sample size to garner trust in the overall system functionality.

\subsection{$\eta$-Threshold Selection}
High $\eta$ values can lead to a very conservative thresholding for the different PPCs in the AAAD, while low $\eta$ values can lead to
aggressive thresholding in the multiple PPCs. Thus, finding an ``optimal'' value for $\eta$ requires fine-tuning. 
To allow for this, we ran two preliminary experiments
with the AAAD, (one aggressive, one conservative), to later interpolate a value that seemed reasonable, $\eta=0.025$. Note that an aggressive $\eta$ might 
lead to observers ignoring the AAAD, and a conservative $\eta$ could be practically irrelevant to implement given its low efficiency benefits.

The optimal $\eta$ value will also depend on the nature of the stimuli, the rigor of the task, and the level of expertise of the participants. For example: pilots, radiologists, security scanning personnel \emph{vs} children,
undergraduates, naive observers.

\subsection{Exploration Map Use}
The Exploration map was rarely used by the observers. Two observers did not use it at all, and one observer used the map on average at least once per trial.
The mean number of times the exploration map was used per trial across all observers was $0.20\pm0.09$ requests$/$trial. 
Notice that observers observers can request the Exploration map more than once per trial.

While we were not rigorous on participant feedback, more than half the users informally reported 
\textit{``I did not find the Exploration map useful, if anything I found it confusing''}, less than half of the users 
reported \textit{``I used it whenever I couldn't find the target and to double-check my decisions''}. This 
response might be due to the short display time (120ms), which might
be insufficient for visually processing the map. 


\subsection{Number of Trials Comparison}
While we found that there is significance in the number of trials accomplished between the two conditions of Experiment 2, there are other factors why on average there might not be a greater difference
across all participants (proportional to say the mean trial time difference). A possible reason is that some participants had smooth runs during Experiment 2 with very little or 
few broken 
fixations during the first stage of both conditions (See Figure~\ref{fig:AAAD_Flow}), while other participants had more broken fixations in one 
Experiment or the other. Pre-trial broken fixations can be due to a subject wearing glasses, eye shape, iris color, pupil size, ethnicity, poor 
initial calibration, \textit{etc.}. These are external factors that can't be controlled for,
and is also why we also emphasize the significance of our results for the average trial time across subjects, which is independent of how many broken fixations they had prior
to each trial, or how many trials they have accomplished. 

Furthermore we performed two additional related samples
t-tests to check if there were any differences in terms of response time for both tasks (target detection and classification), 
but did not find such differences:($M_P=0.12,SD_P=0.19,t_P(17)=1.161,p=0.262$,two-tailed;$M_W=-0.12,SD_W=0.35,t_W(17)=-1.771,p=0.094$,two-tailed).

\subsection{Participant Feedback}
More than one participant, informally reported \textit{``I felt like the AAAD did not help me''}, as well as \textit{``The AAAD helped me confirm my decisions''} and 
both of these opinions seemed
to be spread out across the pool of participants, and did not seem to hold any relationship with their actual performance. 
Our most interesting feedback was given by 
two or three participants who explicitly mentioned that they felt like the AAAD was indirectly \emph{pressuring} them to complete each trial before it fired on. This last feedback is quite
interesting, since it implies that behavior for certain individuals was motivated by trying to \emph{beat} the AAAD, rather than seeing it as a complimentary 
aid to for search. This should be explored in future work.

\end{document}